\documentclass[
	aps,prd,
	reprint,
	showkeys,
	amsmath,amssymb,
	nofootinbib,
	floatfix
	]{revtex4-2}
	
\usepackage[utf8]{inputenc}
\usepackage{fontenc}
\usepackage{amsmath}
\usepackage{aas_macros}
\usepackage{threeparttable}
\usepackage{graphicx}
\usepackage{xcolor}
\usepackage{hyperref}
\hypersetup{
			colorlinks=true,
			pdfauthor={Ebrahim Hassani},
			pdftitle={Is Dark Matter Responsible for the Multiple Stellar Populations in Globular Clusters?},
			citecolor=blue,
			urlcolor=blue, 
			linkcolor=red}

\begin{document}
\title{Is Dark Matter Responsible for the Multiple Stellar Populations in Globular Clusters?}
\author{Ebrahim Hassani}
\email{ebrahim.hassani@birjand.ac.ir, eb.hassani7@gmail.com}
\affiliation{Faculty of Physics, University of Birjand, Birjand, Iran}
\author{Seyyed Milad Ghaffarpour Mousavi}
\email{miladmousavi.96@ms.tabrizu.ac.ir, Miladmousavi5@gmail.com}
\affiliation{Faculty of Physics, University of Tabriz, Tabriz, Iran}

\date{\today}

\begin{abstract}
According to the traditional view of globular clusters, stars inside globular clusters evolved from the same giant molecular cloud. Then the initial chemical compositions of all-stars inside globular clusters must be the same. But recent photometric and spectroscopic observations of globular clusters reveal the presence of more-than-one stellar populations (and with different chemical compositions) inside globular clusters. These findings challenge our traditional view of globular clusters. \\
In this work, the possibility of solving multiple stellar populations problems in globular clusters using dark matter (DM) assumptions is investigated. The presence of DM inside globular clusters changes the physical parameters (e.g. chemical composition, luminosity, temperature, age, etc.) of stars inside them. The distribution of dark matter inside globular clusters is supposed to be non-uniform. It means stars in high dark matter density environments (like the central region of globular clusters) are more affected by the presence of dark matter. Using DM assumption, we showed stars in different locations of globular clusters (corresponding to different dark matter densities that surrounds the stars) follow different evolutionary paths (e.g. on Hertzsprung-Russell diagram). Using the results of simulated stars in different DM density environments, we showed that DM can be considered as one of the reasons for the presence of multiple stellar populations inside globular clusters.
\end{abstract}


\maketitle


\section{Introduction} \label{Sec: Introduction}
\begin{figure*}
	\centering
	\includegraphics[width=1.6 \columnwidth]{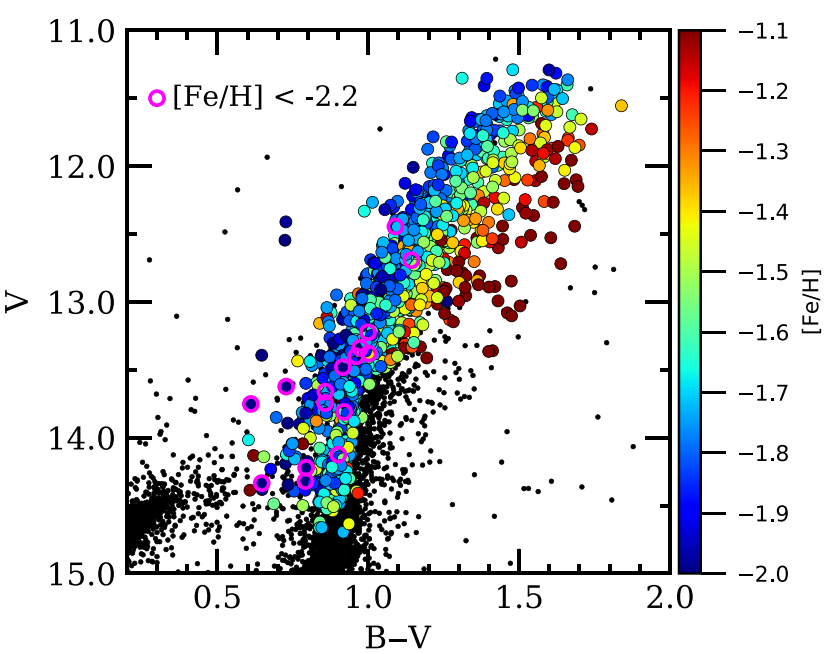}
	\caption{\label{Fig: Color_Magnitude_Diagram_NGC2808} (Colour online) The color-magnitude diagram of $ \omega $ Centauri GC (or NGC 5139, which is the largest known GC of our Galaxy) which depicts the presence of multiple stellar populations. Stars with different [Fe/H] values are color-coded. At least three separate generations of stars are detected in this GC \cite{Gratton2011}. Figure from Ref. \cite{Johnson2020}.}
\end{figure*}

Studying rotation curves of spiral galaxies during the 1960s-1980s leads to the conclusion that about 80 percent of the total mass of galaxies is in the form of dark matter (DM) \cite{1970ApJ...159..379R, Freeman1970,1980ApJ...238..471R}. Subsequent series of studies used different methods (e.g. gravitational lensing \cite{Natarajan2017,Cho2017}, cosmic microwave background analysis \cite{PlanckCollaboration2015}, Lyman-alpha forest \cite{Viel2009,Garzilli2019,Baur2016}, N-body simulations of the universe and galaxies \cite{Vogelsberger2020}, and etc.) to infer that DM must be present in large-scale and small-scale structures (e.g. dwarf galaxies) of the universe. \\
From DM density profiles of galaxies (e.g. NFW or Einasto DM density profile), it is conceivable that DM distributed non-uniformly throughout the galaxies. It means in central regions of galaxies, DM density is the highest and by moving toward the edges of the galaxies, the density of DM reduces. Then it is logical to suppose that all astronomical objects inside galaxies, including stars and stellar clusters, are immersed inside DM. Thus, DM will affect the physics of everything inside galaxies. \\
By passing the time, stars inside galaxies can absorb DM particles from their surrounding. It is predicted that the gathered DM particles inside stars can affect the physics of stars mostly in two ways:
\begin{itemize}
\item  $\: \:$ DM particles can transfer (whether they annihilate or not) energy between different layers of stars. In this way, the temperature profile, pressure profile, chemical composition, and many other physical parameters of stars could be altered in comparison to the standard stellar evolutionary model.
\item $\: \:$ If DM particles annihilate inside stars, then they can act as a new source of energy inside stars, besides the energy that comes from baryonic matter energy production cycles, e.g. pp, CNO and triple $\alpha$ cycles. If this supposition is true, then DM can alter the luminosity and temperature of stars on the H-R diagram too. In this way, stars with the same mass and with the same initial chemical compositions, but with different DM densities that surround them, will follow different evolutionary paths on the H-R diagram (using this note, we could infer that the non-uniform distribution of DM inside GCs can be considered as one of the reasons for the presence of multiple stellar populations in GCs (see Sec (\ref{Sec: Results and simulations}) for more details. In addition, for detailed review about DM effects on stars see the papers: \cite{TurckChieze:2012dc,2009MNRAS.394...82S}).
\end{itemize}
In the last few-decades signs of DM effects on stars were investigated in the literature. For instance:
\begin{itemize}
\item In simulated dwarf galaxies, DM halo around dwarf galaxies heated up by the stars inside them. Then, the more the dwarf galaxy evolves, the more the DM halo heats up by the stars inside them \cite{Read2019}.
\item Stars near the Galactic massive black hole show signs of young and old stars, simultaneously (a problem known as the "paradox of youth" in the community). Considering DM effects on stars it is possible to solve this problem \cite{Hassani_2020zvz}.
\item In 1978 Steigman proposed DM effects on the sun as a possible solution to the solar neutrino problem \cite{Hassani_2020zvz} (though, after discovering the neutrino oscillations by the Super-Kamiokande experiment in 1998 \cite{1998PhRvL..81.1562F} this problem has been considered as a solved problem by the neutrino oscillations assumption, instead of the DM assumption \cite{Fisher1999}).
\item In addition to normal stars, DM effects on other celestial bodies like the moon \cite{2020PhRvD.102b3024C, 2020PhLB..80435403G}, planets \cite{Leane2021, 2012JCAP...07..046H}, neutron stars (NS) \cite{Rezaei_2017, REZAEI20181, 2018JCAP...09..018B, 2018JHEP...11..096K, 2018ApJ...863..157C, 2013PhRvD..87l3507B, Raj2018, Joglekar2020, Baryakhtar2017}, white dwarf stars (WD) \cite{2019JCAP...08..018D, 2018PhRvD..98f3002C, 2018PhRvD..98k5027G, 2016MNRAS.459..695A}, black holes \cite{2012PhRvD..85b3519M, 2009JCAP...08..024U, Belotsky2014} and binary star systems \cite{Hassani2020b} have been investigated in the literature.
\end{itemize}
Globular clusters (GC) are one of the oldest members of our galaxy, the milky way \cite{VandenBerg2013}. According to the traditional view of GCs, it is believed that \cite{Gratton2019}:
\begin{itemize}
\item Stars inside GCs have a similar chemical composition as they evolved from the same giant molecular cloud.
\item Stars inside GCs have nearly the same age.
\item Stars inside GCs are located from nearly the same distance from the Earth.
\end{itemize}
But nova days spectroscopic and photometric studies of GCs do not accept this traditional view for GCs. If this classical views for GCs is correct, then H-R diagrams of typical GCs are anticipated to be a narrow path. But instead, most of the H-R diagrams of GCs are split at least into two separate paths. As an example, Fig. (\ref{Fig: Color_Magnitude_Diagram_NGC2808}) illustrates the color-magnitude diagram of the $ \omega $ Centauri GC. In this figure, stars with higher values of metallicity are less luminous than the low-metallicity stars. This causes the H-R diagram of the $ \omega $ Centauri GC to be thicker than what is anticipated according to the traditional view of globular clusters. The presence of multiple stellar populations can be detected from spectroscopic analysis of the globular clusters too \cite{Masseron2019, Gratton2011, Wang2020}. As an example, the elemental abundances of stars of the $ \omega $ Centauri GC (or NGC 5139, which is the largest known GC of our Galaxy too), vary from star to star \cite{Masseron2019, Gratton2011}.\\
These findings, and many more, are challenging our current view of stellar and stellar clusters evolutionary models. Many models have been developed to solve the discrepancies between the observed and predicted theoretical features of GCs. But each model has its own deficiency and no model has been able to solve the multiple stellar populations problem (MSPP) completely. Figure 6 of Ref. \cite{Bastian2018a} summarized a list of some famous models with their successes and shortcomings that are developed to solve the MSPP in GCs. \\
As an example, one of the first models that are proposed to solve the MSPP is the "asymptotic giant branch stars (or AGB stars) model" \cite{Cottrell1981}. According to this model, first-generation (1G) stars are formed inside the GC's initial giant molecular cloud. High-mass stars are evolved faster than low-mass ones. When high-mass stars reach the AGB phase, they usually become long-period variable stars and lose mass due to the stellar winds. The ejected material from these high-mass stars will pollute the interstellar medium, where second-generation of stars (2G) are forming. In this way, the initial chemical composition of 1G and 2G stars are not the same, although they are formed inside the same giant molecular cloud \cite{Bastian2018a}. AGB scenarios could solve some aspects of MSPP successfully, but it has shortcomings too. For instance, the fraction of 1G and 2G stars are almost the same inside most of the observed GCs. But AGB scenario predicts that just a small fraction of 2G stars will be produced by this way (a problem known as the "mass-budget problem" in the community). In addition, the AGB scenario can not produce observed Na-O anti-correlation which is seen in the spectra of many GCs that host multiple stellar populations \cite{Bastian2018a,Conroy2011}.\\
Another example of the scenarios that are proposed to solve the MSPP is the "fast-rotating massive stars (FRMSs) scenario" \cite{Decressin2007, Decressin2007a}. High-mass stars burn Hydrogen in central regions and through the CNO energy production cycle. The by-products elements of the CNO cycle are different from the pp cycle \cite{2012sse..book.....K}. Fast rotation of high-mass stars, brings the enriched material from central regions to the outer regions of the stars. In the next, the ejected material from FRMSs will pollute the interstellar medium of the host GCs. Similar to the AGB scenario, the 2G of stars will form and evolve by the polluted material that is produced by the 1G FRMSs. Like the AGB scenario, FRMSs scenario, besides its successes, suffers from the mass-budget problem \cite{Bastian2018a, Conroy2011}.
In the current study, we investigated the possibility of solving MSPP using DM assumptions. We used the previously predicted DM effects on stars to provide an alternative possible solution to the MSPP. \\
In Sec. (\ref{Sec: DM effects on stars}), as mentioned above, we reviewed DM effects on stars. Sec. (\ref{Sec: DM in globular clusters}) devoted to discussion on the DM content of the GCs and DM density distribution inside GCs. The results of our simulations are presented in Sec. (\ref{Sec: Results and simulations}). Finally, a discussion about our simulations and results is presented in Sec. (\ref{Sec: Discussions}).
\section{Dark Matter effects on stars} \label{Sec: DM effects on stars}
By passing the time, stars inside galaxies can absorb and accrete DM particles. According to the definition, capture rate of DM particles by a massive body (like stars, planets, white dwarfs, etc.) is the total number of DM particles that are (after weak interaction with baryonic matters) absorbed and gravitationally bounded by that massive body and in units of time \cite{Gould1987}. The capture of DM particles by Hydrogen atoms inside stars can be calculated using Eq. (11) of the paper \cite{Hassani2020b}:
\begin{multline} \label{Eq: CR_by_Hydrogen_atoms}
C_{\chi ,H} = \left [ 4\sqrt{6\pi } \frac{\rho_{\chi}}{m_{\chi}} \frac{1}{\overline{v}_{\chi }v_{\ast}} exp(-\frac{3v^{2}_{\ast}}{2\overline{v}^{2}_{\chi}}) \right ] \\
\left [ \sigma_{\chi,SI} + \sigma_{\chi,SD} \right ] 
\left [ \int_{0}^{R_{\ast}} n_{H}(r) r^{2} dr \right ] \times \\
[ \int_{0}^{\infty } exp(-\frac{3u^{2}}{2\overline{v}^{2}_{\chi}}) sinh(\frac{3uv_{\ast}}{\overline{v}^{2}_{\chi}}) (v_{e}^{2}-\frac{\mu_{-,H}^{2}}{\mu_{H}}u^{2}) \theta (v_{e}^{2}-\frac{\mu_{-,H}^{2}}{\mu_{H}}u^{2}) \\ du ].
\end{multline}
Capture of DM particles by elements heavier than Hydrogen can be calculated using the Eq. (12) of the same paper \cite{Hassani2020b}:
\begin{multline} \label{Eq: CR_by_heavier_elements}
C_{\chi ,i} = \left [ 8\sqrt{6\pi } \frac{\rho_{\chi}}{m_{\chi}^{2}} \frac{E_{0}}{\overline{v}_{\chi }v_{\ast}} \frac{\mu^{2}_{+,i}}{\mu_{i}} exp(-\frac{3v^{2}_{\ast}}{2\overline{v}^{2}_{\chi}}) \right ] \\
\left [ \sigma_{\chi,SI} A_{i}^{2} (\frac{m_{\chi}m_{n,i}}{m_{\chi}+m_{n,i}})^{2}(\frac{m_{\chi}+m_{p}}{m_{\chi}m_{p}})^{2} \right ]
\left [ \int_{0}^{R_{\ast}} n_{i}(r) r^{2} dr \right ] \times \\
[ \int_{0}^{\infty } exp(-\frac{3u^{2}}{2\overline{v}^{2}_{\chi}}) sinh(\frac{3uv_{\ast}}{\overline{v}^{2}_{\chi}}) \: \times  \\
  \left \{ exp(-\frac{m_{\chi}u^{2}}{2E_{0}}) - exp(-\frac{m_{\chi}u^{2}}{2E_{0}}\frac{\mu_{i}}{\mu^{2}_{+,i}}) exp(-\frac{m_{\chi}v_{e}^{2}}{2E_{0}}\frac{\mu_{i}}{\mu^{2}_{-,i}} (1-\frac{\mu_{i}}{\mu^{2}_{+,i}})) \right \} \\
  du  ].
\end{multline}
In Eqs. (\ref{Eq: CR_by_Hydrogen_atoms}) and (\ref{Eq: CR_by_heavier_elements}), 
$ \rho_{\chi} $ is DM density in the location of stars, 
$ m_{\chi} $ is the mass of DM particles, 
$ m_{n,i} $ is the nuclear mass of the element i, 
$ m_{p} $ is the mass of a proton, 
$ A_{i} $ is the atomic number of the element i, 
$ \overline{v}_{\chi} $ is the velocity dispersion of DM particles, 
$ v_{\ast} $ is the velocity of the star relative to the DM halo, 
$ u $ is the velocity of DM particles (velocity distribution of DM particles in the location of stars is usually considered to be a Maxwell–Boltzmanian distribution \cite{2009MNRAS.394...82S} ), 
$ v_{e} $ is the escape velocity from the surface of stars, 
$ \sigma_{\chi,SD} $ is the spin-dependent DM-nucleon scattering cross-section, 
$ \sigma_{\chi,SI} $ is the spin-independent DM-nucleon scattering cross-section, 
$ n_{H}(r) $ is the number density of the Hydrogen atoms at distance r from the centre of the star, 
$ n_{i}(r) $ is the number density of the element i at distance r from the centre of the star, 
$ R_{\ast} $ is the radius of the star, 
$ \theta $ is the step function, 
$ E_{0} = (3 \hbar^{2})/(2 m_{n,i}(0.91 m_{n,i}^{1/3}+0.3)^{2})$ is characteristic coherence energy (see Ref. \cite{Gould1987} for more details) and is a constant, 
$ \mu_{i}$ and $\mu_{\mp,i}$ are defined to be:
$ \mu_{i} = m_{\chi}/m_{n,i} $ , $ \mu_{\mp,i} = (\mu_{i}\mp1)/2 $. \\
Each of Eqs. (\ref{Eq: CR_by_Hydrogen_atoms}) and (\ref{Eq: CR_by_heavier_elements}) consisted of four brackets. The first two brackets are constant and can be calculated analytically. The third brackets are functions of the distance from the centres of stars, $ r $. Although it seems impossible to calculate the third brackets analytically, but it is possible to calculate it numerically using the state-of-the-art stellar evolutionary codes. In this study, we embedded Eqs. (\ref{Eq: CR_by_Hydrogen_atoms}) and (\ref{Eq: CR_by_heavier_elements}) in the last version (version MESA-r21.12.1) of the MESA stellar evolutionary code to calculate the capture rate of DM particles by stars. MESA is a publicly available and open-source code and can simulate the evolution of stars from very low-mass to very high-mass ones ($ 10^{-3} - 10^{3} \: M_{\odot} $). For full capabilities of the MESA code see its official papers \cite{MESA_2015,2013ApJS..208....4P,2011ApJS..192....3P,MESA_2019,MESA_2018}. \\
If DM particles will annihilate inside stars then, they can act as a new source of energy inside stars (beside the energy sources that come from pp and CNO baryonic matter energy production cycles). By multiplying Eqs. (\ref{Eq: CR_by_Hydrogen_atoms}) and (\ref{Eq: CR_by_heavier_elements}) by the $ m_{\chi} c^{2} $ it is possible to calculate the extra luminosity that is produced by DM particles annihilation:
\begin{equation} \label{Eq: DM_particles_annihilation}
L_{\chi} = (Capture \: rate) \times m_{\chi} c^{2},
\end{equation} 
which $ L_{\chi} $ is the luminosity that is produced through DM particles annihilation, and $ c $ is the speed of light.
\section{DM in globular clusters} \label{Sec: DM in globular clusters}
In this section, we want to make an order-of-magnitude estimate of the average DM density inside GCs. Although it is believed that most of the DM content of the GCs has been stripped away by the tidal interactions with their host galaxies, but a typical GC can still keep about 20 percent of its initial DM content \cite{Baumgardt2008}. \\
To estimate the average DM density of a typical GC, consider $ \omega $ Centauri GC (or NGC 5139 globular cluster) as an example. Its physical parameters are presented in table (\ref{Tab: physical_parameters_of_GCs_in_the_milky_way}). According to table (\ref{Tab: physical_parameters_of_GCs_in_the_milky_way}), the total mass of this GC is about $ M = 3.34 \times 10^{6} \: M_{\odot} $ and its V-band mass-to-light ratio is about $ \Upsilon $ = 2.68 $ (M_{\odot}/L_{\odot}) $. Being $ \Upsilon $ bigger than one ($ \Upsilon > 1$) means that we do not receive any light from about 63 percent (that is $ (2.68-1)/2.68 \times 100 = 63 \% $ ) of the total mass of $ \omega $ Centauri GC. Assuming that about 20 percent of the total mass of the $ \omega $ Centauri GC is in the form of DM (and the rest of the dark mass is in the form of white dwarfs, neutron stars, interstellar gas, etc.), then the total DM mass of this GC becomes: \\
\begin{equation} \label{Eq: DM_Mass_of_w_centauri}
M_{\chi, \omega \: Centauri} = \dfrac{20}{100} \times 3.34 \times 10^{6} M_{\odot} = 0.67 \times 10^{6} M_{\odot}.
\end{equation} 
Then, the average DM density inside $ \omega $ Centauri GC becomes:
\begin{equation} \label{Eq: DM_Mass_of_w_centauri}
\overline{\rho}_{\chi, \omega \: Centauri} = \dfrac{\dfrac{1}{2} M_{\chi, \omega \: Centauri}}{\dfrac{4}{3} \pi R^{3}} = 70 \: (M_{\odot}/pc^{3}).
\end{equation} 
In Eq. (\ref{Eq: DM_Mass_of_w_centauri}), R is the half-light radius of the $ \omega $ Centauri GC. We assumed that half of the total mass of this GC lies within the half-light radius. For this reason, the numerator of the Eq. (\ref{Eq: DM_Mass_of_w_centauri}) is multiplied to $\dfrac{1}{2}$. According to Eqs. (\ref{Eq: DM_Mass_of_w_centauri}) and (\ref{Eq: DM_Density_Sun}), the average DM density inside $ \omega $ Centauri GC is about $ 7 \times 10^{3} $ times more than the DM density at the sun's location. DM density at the sun's location was estimated to be about \cite{Salucci2010}:  
\begin{equation} \label{Eq: DM_Density_Sun}
\overline{\rho}_{\chi, \odot} = 0.43 \: (GeV/cm^3 ) \simeq 0.01 \: (M_{\odot}/pc^{3}).
\end{equation}
Assuming that about 20 percent of the total mass of the milky way's GCs is in the form of DM, the average DM density for several other GCs is calculated and presented in table (\ref{Tab: physical_parameters_of_GCs_in_the_milky_way}).
\begin{table*}
\centering
\caption{Physical parameters of some known GCs in the milky way galaxy. Data are from the on-line catalog of GCs parameters which is publicly and freely available at: \url{https://people.smp.uq.edu.au/HolgerBaumgardt/globular/parameter.html}. The catalog's reference papers are: \cite{Baumgardt2018,Baumgardt2020b,Baumgardt2017}}
\begin{threeparttable}
\begin{tabular}{|| c | c | c | c | c | c | c | c || }
\hline \hline
\shortstack{ Name } & \shortstack{ Distance \tnote{*} \\ (K \: pc) } & \shortstack{ Radius \tnote{**} \\ (pc) }  & \shortstack{ Mass \tnote{***} \\ $(M_{\odot})$ } & \shortstack{$(M/L)_{V}$ \\ $ (M_{\odot}/L_{\odot}) $} & \shortstack{ $\overline{\rho}_{\chi}$ \tnote{****} \\ $(M_{\odot}/pc^{3})$ } & \shortstack{ $\overline{\rho}_{\chi}$ \tnote{*****} \\ $(\overline{\rho}_{\odot})$ }   \\ \hline

$ \omega $ Centauri  &   5.24  &  10.36 & $ 3.34 \times 10^{6} \: M_{\odot} $  & 2.68 & 70 & $ 7 \times 10^{3} $  \\

NGC 6535  &  6.5   & 3.65 & $ 1.31 \times 10^{4} \: M_{\odot} $  & 3.93 & $ 6.4 $ & $ 6.4 \times 10^{2} $  \\

NGC 6121   &  1.93  & 3.69  & $ 9.3 \times 10^{4} \: M_{\odot} $  & 2.02 & $ 44 $ & $ 4.4 \times 10^{3} $  \\

NGC 5466  &  16.0  & 14.03 & $ 5.47 \times 10^{4} \: M_{\odot} $  & 1.52 & $ 0.47 $ & $ 4.7 \times 10 $  \\

NGC 6642  &  8.05  & 1.51 & $ 6.45 \times 10^{4} \: M_{\odot} $  & 2.79 & $ 447 $ & $ 4.47 \times 10^{4} $  \\

NGC 6316  &  11.6  & 4.77 & $ 5.09 \times 10^{5} \: M_{\odot} $  & 2.85 & $ 112 $ & $ 1.1 \times 10^{4} $  \\

NGC 3201  &  4.6   & 6.78 & $ 1.41 \times 10^{5} \: M_{\odot} $  & 2.37 & $ 11 $ & $ 1.1 \times 10^{3} $  \\

NGC 1851  &  11.33  & 2.90 & $ 2.81 \times 10^{5} \: M_{\odot} $  & 1.91 & $ 2.7 $ & $ 2.7 \times 10^{2} $  \\

Ter 4  &  6.7  & 6.06 &  $ 7.95 \times 10^{4} \: M_{\odot} $  & 15.34 & $ 8.5 $ & $ 8.5 \times 10^{2} $  \\

NGC 5466  &  16.0  & 14.03 & $ 5.47 \times 10^{4} \: M_{\odot} $  & 1.52 & $ 0.47 $ & $ 47 $  \\ \hline

\end{tabular}
\begin{tablenotes}
\item[*] Distance from the sun
\item[**] Half-mass radius of the GCs
\item[***] Total mass of the GCs
\item[****] Estimated average DM density of the GCs
\item[*****] Estimated average DM density of the GCs in units of the average DM density in the sun's location. DM density in the sun's location was estimated to be about: $ \overline{\rho}_{\chi, \odot} \simeq 0.01 \: (M_{\odot}/pc^{3} $, \cite{Salucci2010})
\end{tablenotes}
\end{threeparttable}
\label{Tab: physical_parameters_of_GCs_in_the_milky_way}
\end{table*}
The mass-to-light ratio of the central regions of the $ \omega $ Centauri GC was estimated to be about $ (M/L)_V = 6.7 (M_{\odot}/L_{\odot}) $ which is higher than the average mass-to-light ratio of the whole GC (\cite{Watkins2013}). This means DM density in central regions of this GC is higher than the average DM density of the whole GC. \\
If DM affects the physics of stars inside GCs, then the evolutionary courses of stars must deviate from the standard stellar evolutionary model. The more the DM density that surrounds the stars is, then the more the deviations from the standard stellar evolutionary model must be. In section (\ref{Sec: Results and simulations}), the results of our simulations showed that stars with the same mass and with the same initial chemical compositions but in different DM density environments (corresponding to different locations inside GCs) follow different evolutionary paths on the H-R diagram. We used these results to propose DM effects on stars as a possible solution for the MSPP in GCs. \\
The calculations of this section aim to show that the average DM density inside Milky-Way's GC is usually several orders of magnitude higher than the average DM density that surrounds the sun. So, DM effects on stars inside GCs are significant and must be taken into account. In the rest of this study, we supposed DM distributed non-uniformly inside GCs, i.e. its density is higher in central regions of the GCs. So, stars in different locations of GCs and with the same mass and with the same chemical composition will follow different evolutionary paths on the H-R diagram.
\section{Results of simulations} \label{Sec: Results and simulations}
\begin{figure*}
	\centering
	\includegraphics[width=2 \columnwidth]{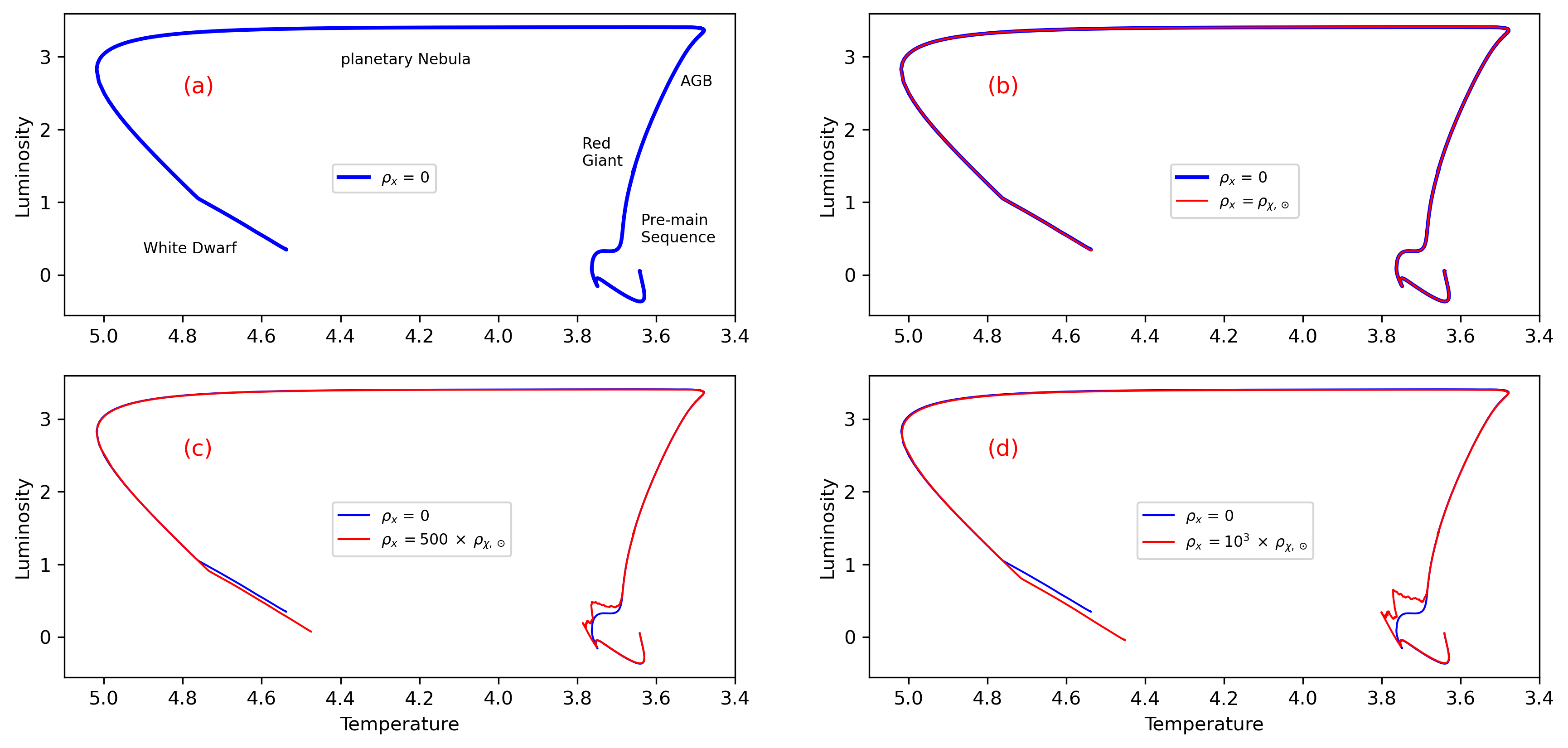}
	\caption{\label{Fig: H_R_Diagram_1_Msun} (Colour online) The evolutionary path of a one-solar-mass star on the H-R diagram and in different DM density environments. In all sub-plots, blue lines represent the evolutionary path of a one-solar-mass star and according to the standard model stellar evolutionary model ($ \rho_{\chi} = 0 $). So, blue lines are the same in all sub-plots. Red lines represent the evolutionary path of a one-solar-mass star when DM effects are considered into account ($ \rho_{\chi} \neq 0 $). From sub-plot \ref{Fig: H_R_Diagram_1_Msun}-b to sub-plot \ref{Fig: H_R_Diagram_1_Msun}-d the density of DM increased. So, from sub-plot \ref{Fig: H_R_Diagram_1_Msun}-b to sub-plot \ref{Fig: H_R_Diagram_1_Msun}-d, the deviation between blue and red lines increased too.}
\end{figure*}

In our simulations, weakly interacting massive particles (WIMP) with masses $ m_{\chi} = 100 \: Gev \: c^{-2} $ considered to be the DM candidate. In Eqs. (\ref{Eq: CR_by_Hydrogen_atoms}) and (\ref{Eq: CR_by_heavier_elements}), spin-dependent and spin-independent scattering cross-sections considered to be $ \sigma_{\chi,SD} = 10^{-38} \: cm^{2}$  and $ \sigma_{\chi,SI} = 10^{-44} \: cm^{2} $, respectively. These amounts are the maximum magnitudes that are determined through the experimental DM detection experiments \cite{2008PhRvL.100b1303A, 2008PhRvL.101i1301A}. Escape velocity from the surface of the stars $v_{e}$ are calculated while each star is evolving. We used MESA's build-in functions to calculate $v_{e}$ in each time-step of the evolution of stars. In addition, the velocity distribution of dark matter particles is assumed to be a Maxwell-Boltzmann distribution with a velocity dispersion $ \overline{v}_{\chi} = 270 \: km.sec^{-1}$ \cite{2009MNRAS.394...82S}. The velocity of stars relative to the dark matter halo of the GCs is considered to be $ v_{\ast} = 20 \: Km.sec^{-1} $. \\
After considering Eq. (\ref{Eq: DM_particles_annihilation}) in MESA stellar evolutionary code and running it, the results of the simulations for a one-solar-mass star are presented in Fig. (\ref{Fig: H_R_Diagram_1_Msun}). In this figure, blue lines represent the evolutionary paths of a one solar mass star according to the standard stellar evolutionary model (i.e. when DM effects have not been considered into account). So, the blue lines are the same for all sub-plots in Fig. (\ref{Fig: H_R_Diagram_1_Msun}).
\begin{figure*}
	\centering
	\includegraphics[width=2 \columnwidth]{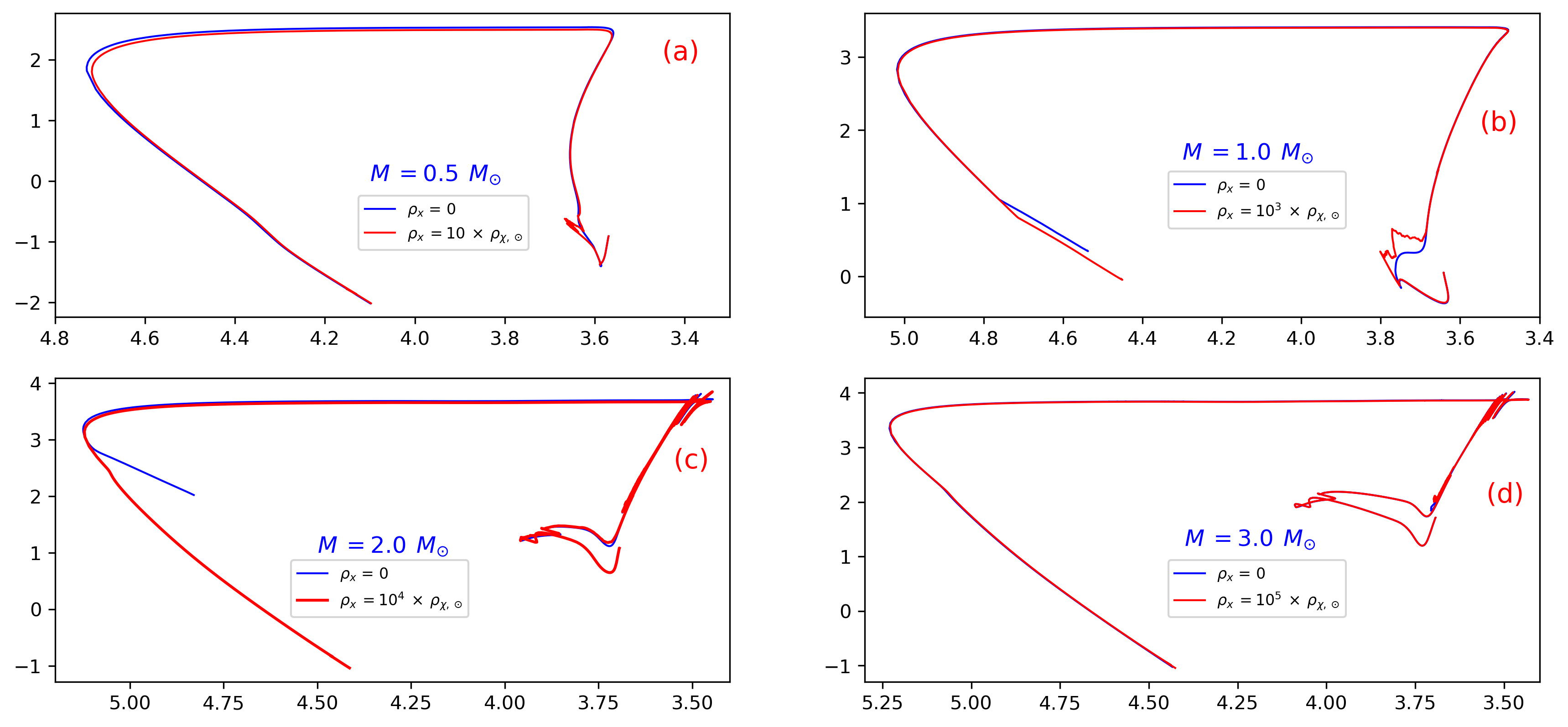}
	\caption{\label{Fig: H_R_Diagram_WIMPy_Stars} (Colour online) The evolutionary paths of stars with different masses on the H-R diagram. Like Fig. (\ref{Fig: H_R_Diagram_1_Msun}), blue lines represent the evolutionary paths of stars according to the standard stellar evolutionary model ($ \rho_{\chi} = 0 $). Red lines represent the evolutionary paths of stars when DM effects are taken into account ($ \rho_{\chi} \neq 0 $). Sub-plots  \ref{Fig: H_R_Diagram_WIMPy_Stars}-a, \ref{Fig: H_R_Diagram_WIMPy_Stars}-b, \ref{Fig: H_R_Diagram_WIMPy_Stars}-c and \ref{Fig: H_R_Diagram_WIMPy_Stars}-d represent the evolutions of stars with masses $0.5 \: M_{\odot}$, $ 1.0 \: M_{\odot}$, $ 2.0 \: M_{\odot} $ and $ 3.0 \: M_{\odot} $, respectively. It is conceivable from the figure that, by considering DM effects, the evolutionary paths of stars will deviate from the standard stellar evolutionary model.}
\end{figure*}
Red lines in Fig. (\ref{Fig: H_R_Diagram_1_Msun}) represent the evolutionary paths of one-solar-mass stars when DM effects are taken into account. Each red line in each sub-plot corresponds to the evolutionary path of a star with different DM density that surrounds the star (i.e. different values for $ \rho_{\chi} $ ). In Fig. (\ref{Fig: H_R_Diagram_1_Msun}) by increasing DM density from Fig. (\ref{Fig: H_R_Diagram_1_Msun}-b) to Fig. (\ref{Fig: H_R_Diagram_1_Msun}-d), the deviation between red and blue lines increases. So, we can say, the presence of a non-uniform distribution of DM inside GCs causes stars with the same mass and with the same initial chemical composition but with different values of $ \rho_{\chi} $ to follow different evolutionary paths on the observed H-R diagram of GC (e.g. Fig. (\ref{Fig: Color_Magnitude_Diagram_NGC2808})). \\
Fig. (\ref{Fig: H_R_Diagram_WIMPy_Stars}) represents the results of our simulations for stars with different masses. Like Fig. (\ref{Fig: H_R_Diagram_1_Msun}), blue lines represent the evolution of stars according to the standard stellar evolutionary model. Red lines represent the evolution of stars when DM assumptions are taken into account. It is conceivable from Fig. (\ref{Fig: H_R_Diagram_WIMPy_Stars}) that, when DM effects are taken into account, stars with the same mass (in each sub-plot) and with the same initial chemical compositions follow different evolutionary paths on the H-R diagram. \\
Assuming that DM density distribution inside GCs is not uniform, then stars in different locations of GCs are immersed in different DM density environments. So, stars with the same mass and with the same initial chemical compositions will follow different evolutionary paths on the H-R diagram. These results might be the solution to the question: why we see different generations of stars inside GCs? We will discuss more about this in Sec. (\ref{Sec: Discussions}).
\section{Discussions and Conclusions} \label{Sec: Discussions}
According to the results of our simulations (our results for stellar simulations in the presence of DM are in agreement with the results of previous works, e.g. see (\cite{2009MNRAS.394...82S})), the presence of DM inside GCs causes the evolutionary paths of stars to deviate from the standard stellar evolutionary model. The more the DM density around a star is, then the more the stars deviate from the standard stellar evolutionary model (see Fig. (\ref{Fig: H_R_Diagram_1_Msun}) and Fig. (\ref{Fig: H_R_Diagram_WIMPy_Stars})). \\
It is anticipated that GCs inside galaxies have lost most of their DM content due to the tidal interactions with their host galaxies \cite{Wirth2020}. But in Sec. (\ref{Sec: DM in globular clusters}) we discussed about the possibility that GCs may have kept some portion of their initial DM content. Mass-to-light ratios of most of the Milky way's GCs are bigger than one (see table (\ref{Tab: physical_parameters_of_GCs_in_the_milky_way})). It means we do not receive any light from some mass of the GCs. Assuming that just 20 percent of this dark part of the GCs is in the form of DM, then we estimated the average DM density for some GCs. In almost all cases, $ \overline{\rho}_{\chi} $ is several times bigger than the average DM density around the sun (see table (\ref{Tab: physical_parameters_of_GCs_in_the_milky_way}) for estimated values for $ \overline{\rho}_{\chi} $). \\
Assuming that DM distributed non-uniformly inside GCs (like NFW or Einasto DM density profile \cite{Merritt2006a}), then stars with the same mass and with the same initial chemical compositions will follow different evolutionary paths on the H-R diagram. As an example, a one-solar-mass star near the central regions of a GC (and with the higher DM density that surrounded the star) will follow different evolutionary path on the H-R diagram in comparison to a one-solar-mass star that evolves near the outer regions of the same GC (see Fig. (\ref{Fig: H_R_Diagram_1_Msun})). \\
In addition to the location of stars on the H-R diagram, the presence of DM can alter the amount of time that stars spent in each evolutionary phase (e.g. main-sequence phase or red-giant phase) of their evolution. As an example, consider the main-sequence phase. According to definition, it is a phase at which most of the energy sources of stars come from Hydrogen fusion in the core of stars. Depending on the mass of the stars, stars convert Hydrogen to Helium through pp or CNO energy production cycles \cite{2000itss.bookP}. Both pp and CNO energy production cycles are a strong function of the temperature (that is, $\varepsilon_{pp} \varpropto T^{4} $ and $\varepsilon_{CNO} \varpropto T^{16} $, \cite{2000itss.bookP}). Then, we can say, if DM particles will annihilate inside stars, then they alter the temperature of the core of the stars too. This causes stars to consume Hydrogen atoms at different rates in comparison to the models without DM. So, we infer that the presence of DM affects the elemental abundances of star too.\\
Our overall result from this discussion is that, if the presence of DM can alter the temperature of stars, then it can alter the age, chemical composition, luminosity, and many other physical parameters of stars too. \\
If the presence of DM alters the rate of pp and CNO energy production cycles inside stars, then we can say that, the chemical compositions of stars can be affected by the presence of DM too. But, because of a lack of our knowledge about the exact physical nature of DM, it is hard to discuss more about the exact consequences of DM effects on stars. \\
Our overall result is that, if the presence of DM alters the luminosity, temperature, chemical composition, age, etc. of stars, then its presence can be considered as a possible solution to the multiple stellar populations problem in GCs.
\section{Acknowledgement} \label{Sec_ACKNOWLEDGMENTS}
Special thanks are due to Dr Amin Rezaei Akbarieh From university of Tabriz, Iran, Prof. Kenneth Freeman from Research School of Astronomy and Astrophysics, Australian National University, Prof. Nate Bastian from the University of Liverpool John Moores, England and Dr. Marco Taoso from National Institute of Nuclear Physics (INFN) Turin, Italy for their helpful discussions during the research. Figures of this work are generated using python's visualizations library: matplotlib v3.2.1 \cite{Hunter2007}. This research made use of the python data analysis library, Pandas \cite{McKinney2010}.
\section{Data availability}
The data underlying this article will be shared on reasonable request to the corresponding author

\bibliographystyle{apsrev4-2_16.bst}
\bibliography{references}

\end{document}